\documentclass[aps,pra,twocolumn,floatfix,showpacs,preprintnumbers,amsmath,amssymb,10pt,nofootinbib]{revtex4-1}
\usepackage[dvips]{graphics}

\DeclareMathAlphabet{\mathsfit}{\encodingdefault}{\sfdefault}{m}{sl}
\SetMathAlphabet{\mathsfit}{bold}{\encodingdefault}{\sfdefault}{bx}{sl}

\usepackage{mathptmx}
\usepackage{psfrag}
\usepackage{graphicx}
\usepackage{bm}
\usepackage{amsmath}
\usepackage{trfsigns}
\usepackage{amssymb}
\usepackage[usenames,dvipsnames]{color}
\usepackage{dashrule}
\usepackage[english]{babel}
\usepackage{contour}
\usepackage{upgreek,bm}
\usepackage{color}
\usepackage{array}
\usepackage[colorlinks,bookmarks=false,citecolor=blue,linkcolor=cyan,urlcolor=blue]{hyperref}

\definecolor{dred}{rgb}{.6,.0,0.}
\definecolor{dblue}{rgb}{.0,.0,0.6}

\newcommand{\bra}[1]{\ensuremath{\left\langle#1\right|}}
\newcommand{\ket}[1]{\ensuremath{\left|#1\right\rangle}}
\renewcommand{\vec}[1]{\mathbf{#1}}

\newcommand{\tens}[1]{\mbox{\textsf{\textbf{#1}}}}

\newcommand{\sprod}{\!\cdot\!}

\newcommand{\mi}{\textrm{i}} 
\newcommand{\me}{\mathrm{e}}

\begin{document}

\title{Superradiance from non-ideal initial states -- a quantum trajectory approach}

\author{Sebastian Fuchs$^1$}
\author{Andr\'{a}s Vukics$^2$}
\author{Stefan Yoshi Buhmann$^1$}

\affiliation{$^1$ Physikalisches Institut, Albert-Ludwigs-Universit\"at Freiburg, Hermann-Herder-Stra{\ss}e 3, 79104 Freiburg, Germany\\
$^2$ Institute for Solid State Physics and Optics, Wigner Research Centre for Physics, P.O. Box 49, H-1525 Budapest, Hungary
}

\date{\today}

\begin{abstract}
Collective emission behavior is usually described by the decay dynamics of the completely symmetric Dicke states. To study a more realistic scenario, we investigate alternative initial states inducing a more complex time evolution. Superposition states of the fully inverted Dicke state and the Dicke ground state with unequal mutual weights are studied as examples as well as superradiance stemming from atoms in clusters separated by more than one wavelength. The Monte Carlo wave function method serves as framework to study the dynamics of quantum states, which is determined by quantum jumps on the one hand and continuous evolution dynamics on the other hand. We compare this method with the classical picture of a system of rate equations written for the diagonal components of the density matrix.
\end{abstract}

\maketitle

\section{Introduction}
Since its theoretical prediction by Dicke in 1954 \cite{Dicke:1954} many aspects of superradiance have been studied. The term refers to the collective enhancement of spontaneous emission of an atom if it is part of a dense atomic ensemble, whose extension is much smaller than the radiation wavelength. This phenomenon has been observed experimentally for the first time in an optically pumped hydrogen-fluoride gas in 1973 \cite{Skribanowitz:1973}.

The Hamiltonian describing the atom--light interaction between the dipole moment of the atomic ensemble and the quantized field mode is known as the Dicke Hamiltonian. In rotating-wave approximation it is called Tavis--Cummings Hamiltonian \cite{Tavis:1968}, which is the extension of the Jaynes--Cummings Hamiltonian \cite{Jaynes:1963} for collective spin operators replacing two-level Pauli spin matrices. It was found that Tavis--Cummings and Dicke Hamiltonians show a phase transition in the thermodynamic limit $N \rightarrow \infty$ \cite{Hepp:1973, Wang:1973}. This superradiant phase is characterized by macroscopic occupations in the electromagnetic field and macroscopic excitation in the atoms. Later it was argued that a term quadratic in the electromagnetic vector-potential $\vec{A}^2$ is missing in the Dicke Hamiltonian which makes that the superradiant phase is unphysical \cite{Rzazewski:1975}. This no-go theorem initiated an ongoing debate about the validity of the Dicke model. Recently, the Dicke Hamiltonian in its well-known form was rederived in the minimal and multipolar coupling pictures \cite{Vukics_2:2012, Vukics:2014}. The experimental observation of the superradiant phase transition \cite{Baumann:2010} reinforced the relevance of the Dicke model also for systems very different from Dicke’s original setting. A connection between the dynamical feature of superradiance \cite{Dicke:1954} and the phase transition picture \cite{Hepp:1973, Wang:1973} was established by studying the dynamics of a system of several atoms and a damped cavity \cite{Fuchs:2016}. In this model the superradiant burst becomes apparent as a peak of the cavity excitation and the occupation of the steady state mimics the superradiant phase transition depending on the coupling strength between the atoms and the cavity mode.

In this paper we concentrate on one facet of superradiance: the emission burst. Motivated by recent experiments with excited molecules \cite{Muller:2013, Muller:2015}, where not all emitters can be brought into the excited state, the question arises whether a partially excited ensemble of emitters still emits superradiantly. In general there are several options for a quantum state corresponding to a partially excited atomic ensemble. In Ref.~\cite{Nefedkin:2016} superradiance from non-Dicke states is investigated, where the proportionality of the intensity to the number of atoms squared is considered as indicator for superradiance. Mixed non-Dicke states are considered as initial states. Among other things, an initial density matrix of an atomic ensemble is generated by the product of the single-atom density matrices. This is contrasted to the Dicke model and a connection to the phase operator for two-level atoms is established. Ref. \cite{Manassah:2014} studies a partially inverted or excited slab of two-level atoms. The superradiant emission is calculated in the basis of eigenmodes of the Lienard-Wiechert Green's function. Depending on the initial excitation level of the slab, three regimes are distinguished. If the initial excitation is lower than the number of excitations at the maximum, a superradiant burst is not observed. A weak superradiant burst can be observed if the initial excitation is higher than the one at the maximum but below a certain threshold. Above this threshold the superradiant burst is significant.

Especially Refs.~\cite{Nefedkin:2016, Manassah:2014} show the importance of initial conditions for the observation of a superradiant burst. Here we investigate the dependence of superradiance on different initial conditions, namely a completely inverted ensemble of two-level systems, the classical mixture of the completely inverted state and the deexcited state and the semi-excited Dicke state with half of the atoms in the excited state. Moreover, we do not restrict ourselves to Dicke states only, but study coherent superpositions of the excited and deexcited Dicke states and additionally look at the radiation pattern of an atomic ensemble split into several clusters that are separated by distances surpassing the wavelength of the atomic radiation.

This paper is organized as follows. Two different theoretical models to simulate different initial conditions are discussed in Sec.~\ref{sec:Theoretical model}: Emission rate equations for Dicke states \ref{sec:Emission rate equations with Dicke states} and the formalism of Monte Carlo wave function (MCWF) for a single two-level system \ref{sec:Monte Carlo wave functions}. Section \ref{sec:Comparison of the methods} discusses the application of the MCWF to fully symmetric states (Dicke states), and the dynamics is compared to the classical rate equation approach. The dynamics using several initial conditions, i.e. the coherent superposition of the completely inverted Dicke state and the ground state as well as the semi-inverted completely symmetric Dicke state, is studied in Sec.~\ref{sec:Superradiance for several initial states}. Moreover, superposition states of the completely inverted Dicke state and the ground state of unequal weights are investigated. Finally, we study superradiance with atoms in two separated clusters (\ref{sec:Collective emission from two chunks}).

\section{Theoretical model}
\label{sec:Theoretical model}
The Dicke model developed in Ref.~\cite{Dicke:1954} is outlined in Sec.~\ref{sec:Emission rate equations with Dicke states}, where we introduce the Dicke states and collective emission. Using these concepts, in Sec.~\ref{sec:Simple manifestation of superradiance for 2 atoms} the system of rate equations for the Dicke states is introduced, and the possibility of superradiance is demonstrated in the simplest scenario of $2$ atoms. Afterwards, in Sec.~\ref{sec:Monte Carlo wave functions}, we outline the MCWF method (also known as quantum-jump Monte Carlo).

\subsection{Emission rate equations between Dicke states}
\label{sec:Emission rate equations with Dicke states}
The Dicke model and the concept of superradiance was developed in Dicke's original paper \cite{Dicke:1954}. It studies the collective emission of an ensemble of two-level atoms with an extension of the ensemble which is small compared to the wavelength. In this scenario the atoms can be excited by a laser field and emit into free space. As opposed to cavity quantum electrodynamics (CQED) scenarios, excitations once emitted cannot be reabsorbed by the atoms. There is no mutual exchange of excitations between the atoms and the quantized electromagnetic field described by the Dicke Hamiltonian. Further sources of energy loss such as atomic collisions are not considered in this model. A modern description of this model can be found in Refs.~\cite{Haroche:1982} and \cite{Brandes:2005}.

In particular, the collective emission model adopted here is explained in great detail in Sections 1--3 of Ref.~\cite{Haroche:1982}. It contains two essential ingredients necessary for collective spontaneous emission from an atomic ensemble:
\begin{enumerate}
 \item The ensemble interacts with a continuum of modes (free space scenario as opposed to CQED), each electromagnetic mode of the continuum coupled to the atoms according to the Dicke Hamiltonian. Such a continuum of modes forms a reservoir, and makes that the atomic ensemble (the small subsystem of the large coupled system) behaves in a dissipative way.
 \item Assuming atomic transition frequencies in the optical domain, a Markov approximation can be made on the electromagnetic reservoir. This is because the reservoir correlation time scales with the optical frequency, while the characteristic timescale of the dynamics of the atomic ensemble with the spontaneous emission rate, and the two are typically separated by several orders of magnitude. (E.g. for the D$_2$ line of Rubidium 87, the optical transition frequency is $\approx 2\pi\times384\,\text{THz}$, whereas the spontaneous emission rate is only $\approx 2\pi\times6\,\text{MHz}$.)
\end{enumerate}
These two assumptions lead from the Dicke Hamiltonian description to the collective dissipative model for the atomic ensemble detailed below.

The atomic ensemble consists of $N$ identical two-level atoms each with ground state $\ket{g}$ and an excited state $\ket{e}$ separated by the energy $\hbar \omega$. A single two-level atom is described by Pauli-spin matrices and raising and lowering operators for the $i$th atom are defined as
\begin{equation}
\hat{\sigma}^+_i = \ket{e} \bra{g}; \; \; \hat{\sigma}^-_i = \ket{g} \bra{e}  
\end{equation}
with the diagonal operator (population inversion)
\begin{equation}
\hat{\sigma}^z_i = \frac{1}{2} \left( \ket{e} \bra{e} - \ket{g} \bra{g} \right).
\end{equation}
These operators only act in the $i$th subspace and follow the commutation relations for Pauli-spin matrices
\begin{equation}
\left[ \hat{\sigma}^z_i, \hat{\sigma}^{\pm}_j \right] = \pm \delta_{ij} \hat{\sigma}^{\pm}_i; \; \; \left[ \hat{\sigma}^+_i, \hat{\sigma}^-_j \right] = 2 \delta_{ij} \hat{\sigma}^z_i.
\end{equation}
In an idealized scenario, initially at $t=0$ all $N$ atoms are excited in level $\ket{e}$ and the state of the ensemble reads
\begin{equation}
\ket{\psi \left( t=0 \right)} = \ket{e,e,...,e}.
\label{eq:Initial State}
\end{equation} 
Since the atoms are confined to a volume that is small compared to the wavelength, it is not possible to distinguish one specific atom emitting a photon of wavelength $\lambda$ according to Heisenberg's uncertainty principle. Rather, the emission stems from the entire collection of indistinguishable atoms. Therefore, any quantum state of the atomic system has to be symmetrical with respect to the exchange of any two atoms of the ensemble during the whole time evolution. Here, we will make use of this plausible symmetry argument, even though it might not hold in certain situations \cite{Haroche:1982}.

We introduce the collective symmetrical spin operators
\begin{equation}
\hat{J}^{\pm} = \sum\limits^N_i{\hat{\sigma}^{\pm}}; \; \; \hat{J}^z = \sum\limits^N_i{\hat{\sigma}^z}
\label{eq:Collective Operators}
\end{equation}
and
\begin{equation}
\hat{\vec{J}}^2 = \frac{1}{2} \left( \hat{J}^+ \hat{J}^- + \hat{J}^- \hat{J}^+ \right) + \left( \hat{J}^z \right)^2,
\end{equation}
which follow the commutation relations of angular momentum operators for a spin of modulus $J=N/2$. The states invariant under the permutation of any two atoms are named Dicke states, and are eigenstates $\hat{J}^{z}$. These $N+1$ states can be constructed starting from the maximally excited state \eqref{eq:Initial State} as
\begin{equation}
\ket{J,M} = \sqrt{\frac{\left( J+M \right)!}{N! \left( J-M \right)!}} \left(\hat J^-\right)^{J-M} \ket{e,e,...,e}
\label{eq:Dicke State}
\end{equation}
with $-J \leq M \leq J$. The state $\ket{J,M}$ is fully symmetrical with $J+M$ atoms in the excited state $\ket{e}$ and $J-M$ in the ground state $\ket{g}$.

Acting on the Dicke state \eqref{eq:Dicke State}, the ladder operators $\hat{J}^{\pm}$ fulfill the relations
\begin{align}
\hat{J}^+ \ket{J,M} &= \sqrt{J \left( J+1 \right) - M \left( M+1 \right)} \ket{J,M+1},\nonumber\\
\hat{J}^- \ket{J,M} &= \sqrt{J \left( J+1 \right) - M \left( M-1 \right)} \ket{J,M-1}.
\end{align}
The Dicke state $\ket{J,M}$ is an eigenstate of the operators of collective angular momentum
\begin{equation}
\hat{J}^z \ket{J,M} = M \ket{J,M}; \; \; \hat{\vec{J}}^2 \ket{J,M} = J \left( J+1 \right) \ket{J,M}.
\end{equation}
The $N+1$ collective states $\ket{J,M}$ form an equidistant ladder in energy with level-splitting $\hbar \omega$. Starting with a completely excited atomic ensemble in state \eqref{eq:Initial State}, this system evolves along the ladder of all Dicke states \eqref{eq:Dicke State} down to the collective ground state $\ket{g,g,...,g}$ and thereby emit radiation with time-dependent intensity.

In order to compute the state-dependent collective radiation intensity, the system radiation rate for each state $\ket{J,M}$ with a fixed value of $J$ is needed, which is given by the following quantum mechanical expectation value:
\begin{multline}
\Gamma_{M, M-1} = \Gamma \langle \hat{J}{}^+ \hat{J}^- \rangle_M = \Gamma \bra{J,M}\hat{J}{}^+ \hat{J}^- \ket{J,M} \\= \Gamma \left( J+M \right) \left( J-M+1 \right),
\label{eq:Decay Rates}
\end{multline}
where $\Gamma$ denotes the emission rate of a single atom. It is apparent hence that the collective emission rate starts with a value of $2 J \Gamma$ in the fully excited state \eqref{eq:Initial State} with $M=+J$, where the atoms emit photons independently (this is product state). It reaches its highest value $J \left( J+1 \right) \Gamma$ at $M=0$, proportional to $N^2$, where the collective enhancement of the emission is strongest, and which gives the main contribution to the superradiant burst. Finally, radiation comes to an end at $M=-J$.

The intensity of the radiation $I \left( t \right)$ is the sum of the decay rate $\Gamma_{M,M-1}$ for all Dicke states $\ket{J,M}$ weighed by the time-dependent probability of the system to occupy this state $p_M \left( t \right)$
\begin{equation}
I \left( t \right) = \sum\limits^{J}_{M=-J+1}{p_M \left( t \right) \Gamma_{M, M-1} }.
\label{eq:Emission Intensity}
\end{equation}
The probability distribution obeys the classical master equation
\begin{equation}
\dot{p}_M \left( t \right) = -\Gamma_{M,M-1} p_M \left( t \right) + \Gamma_{M+1,M} p_{M+1} \left( t \right),
\label{eq:System of Rate Equations}
\end{equation}
which can be expressed in matrix-vector representation with a time-dependent probability vector $\vec{p} \left( t \right)$ as
\begin{equation}
\dot{\vec{p}} \left( t \right) = \tens{A} \sprod \vec{p} \left( t \right),
\end{equation}
which can be solved readily:
\begin{equation}
\vec{p} \left( t \right) = \exp \left( \tens{A} t \right) \vec{p} \left( 0 \right).
\label{eq:System of Rate Equations Matrix Representation}
\end{equation}
The matrix $\tens{A}$ consists of the constant decay rates \eqref{eq:Decay Rates}.

The intensity shows a radiation burst, whose maximum scales with $N^2$ and the width of the peak exhibits a $1/N$ behavior. The integrated intensity over the time of emission is a measure for the emitted energy and thus the number of emitted photons. It reflects the total number of photons initially brought into the system, which is equal to the total number of atoms in case of maximally excited two-level systems. Thus the integrated intensity is identical to the value of $N$, if the initial state is the Dicke state \eqref{eq:Initial State}.

\subsection{Simple manifestation of superradiance for 2 atoms}
\label{sec:Simple manifestation of superradiance for 2 atoms}
To get a feeling of the physics we consider, let us take a brief look at the simple example of $2$ atoms. Here, three Dicke states are involved \eqref{eq:Dicke State}
\begin{align}
\ket{1,1} &= \ket{e,e},\nonumber\\
\ket{1,0} &= \frac{1}{\sqrt{2}} \left( \ket{e,g} + \ket{g,e} \right),\nonumber\\
\ket{1,-1} &= \ket{g,g}.
\end{align}
The system of rate equations reads
\begin{equation}
\begin{pmatrix} \dot{p}_1 \left( t \right) \\ \dot{p}_0 \left( t \right) \\ \dot{p}_{-1} \left( t \right) \end{pmatrix} = \begin{pmatrix} -2 \Gamma & 0 & 0 \\ 2 \Gamma & - 2 \Gamma & 0 \\ 0 & 2 \Gamma & 0 \end{pmatrix} \begin{pmatrix} p_1 \left( t \right) \\ p_0 \left( t \right) \\ p_{-1} \left( t \right) \end{pmatrix}, \; p \left( 0 \right) = \begin{pmatrix} 1 \\ 0 \\ 0 \end{pmatrix}.
\end{equation}
The decay rates $\Gamma_{M,M-1}$ are obtained from Eq.~\eqref{eq:Decay Rates}. The system starts from the fully excited state $\ket{1,1}$. The dynamical matrix has eigenvalues $0$ and $2$, where the latter is doubly degenerate leading to the two solutions $\exp(-2 \Gamma t)$ and $\Gamma t \exp(-2 \Gamma t)$. The first solution of the doubly degenerate eigenvalue causes an exponential decay whereas the second one is responsible for a peaked decay pattern. The occupation probabilities under the given initial condition are found analytically:
\begin{equation}
\begin{pmatrix} p_1 \left( t \right)\\ p_0 \left( t \right)\\ p_{-1} \left( t \right) \end{pmatrix} = \begin{pmatrix} \exp(-2 \Gamma t)\\ 2 \Gamma t \exp(-2 \Gamma t)\\ 1 - \left( 1 + 2 \Gamma t \right) \exp(-2 \Gamma t) \end{pmatrix}.
\end{equation}
Summing up probabilities and decay rates according to Eq.~\eqref{eq:Emission Intensity}, the total intensity gives
\begin{equation}
I \left( t \right) = 2 \Gamma \exp(-2 \Gamma t) \left( 1 + 2 \Gamma t \right).
\end{equation}

The basic structure of eigenvalues and solutions of the system of rate equations \eqref{eq:System of Rate Equations} for an atomic ensemble of more than two atoms is similar. There are pairs of doubly degenerate eigenvalues causing the peaked structure of the photon emission $I \left( t \right)$ \eqref{eq:Emission Intensity} as opposed to a superposition of ordinary exponential terms.

In Appendix \ref{sec:DickePaper}, we review in a formalism more accessible to the contemporary reader the most fundamental manifestation of superradiance that occurs with two emitters, which was presented in Dicke’s original paper~\cite{Dicke:1954}.

Collective enhancement due to the identicity of particles is thus already observed in the case of only two contributors.

\subsection{Monte Carlo wave function method}
\label{sec:Monte Carlo wave functions}
The theoretical model described in Sec.~\ref{sec:Emission rate equations with Dicke states} operates on the subspace spanned by Dicke states. Any other collective atomic state with a high number of excitations is isaccessible by applying the Dicke operators. In order to study superradiance with other initial conditions a different approach, which is not restricted to the Hilbert space of Dicke states, is needed. The description of the MCWF approach, cf. Ref.~\cite{Plenio:1998}, is mostly based on Ref.~\cite{Molmer:1993} which coalesces elements of several preworks \cite{Diosi:1985, Javanainen:1986, Dalibard:1992, Carmichael_Book, Dum:1992, Hegerfeldt:1991}.

A small quantum system which is coupled to a reservoir can be described by the standard master-equation approach from quantum optics, cf. Ref.~\cite{Breuer_Book}. Whereas the master-equation deals with density matrices with $N^2$ components, the MCWF unravels this evolution into an ensemble of stochastic quantum trajectories dealing with state vectors with $N$ components. Along with other benefits this scaling behavior is highly attractive for the study of the time evolution of large quantum systems. The gain in computing time is predominantly due to the possibility of parallelization of quantum trajectory realizations.

The master equation of a system with Hamiltonian $\hat{H}$ relies on the Born-Markov approximation, and can be written as
\begin{equation}
\dot{\hat{\rho}} = \frac{\mi}{\hbar} \left[ \hat{\rho}, \hat{H} \right] + \mathcal{L} \left( \hat{\rho} \right)
\end{equation}
with the relaxation (Liouvillean) superoperator for a single decay channel, such as the decay of photons into the reservoir at zero temperature reading
\begin{equation}
\mathcal{L} \left( \hat{\rho} \right) = \Gamma \left( -\frac{1}{2} \hat{C}^+ \hat{C}^- \hat{\rho} - \frac{1}{2} \hat{\rho} \hat{C}^+ \hat{C}^- + \hat{C}^- \hat{\rho} \hat{C}^+ \right),
\label{eq:Relaxation Operator}
\end{equation}
where $\hat{C}^-$ and $\hat{C}^+$ are ladder operators.

The MCWF method evolves the state vector with a non-Hermitian Hamiltonian
\begin{equation}
\hat{H}_{\textrm{nH}} = \hat{H} - \frac{\mi \hbar \Gamma}{2} \hat{C}^+ \hat{C}^-,
\label{eq:Total Hamiltonian}
\end{equation}
(no-jump evolution), which models the fact that an open system is under continuous observation (weak measurement) by its environment, hence continuously leaking information into the environment, even without actual quantum jumps. This no-jump evolution is interrupted by instantaneous quantum jumps, whose probability derives from the norm loss the state vector suffers through the non-unitary evolution. It is non-trivial how often the possibility of a jump has to be probed in a numerical implementation \cite{Kornyik:2019}.

One way to treat this is to break down the MCWF evolution into steps of $\delta t$ (possibly adaptive timesteps), and probe for the possibility of quantum jumps in each timestep. If the system is in the normalized state $\ket{\psi \left( t \right)}$ at time $t$, then at time $t + \delta t$, under the non-unitary evolution to first order in $\delta t$, its state becomes
\begin{equation}
\ket{\psi' \left( t+\delta t \right)} = \left( 1-\frac{\mi \hat{H}_{\textrm{nH}} \delta t}{\hbar} \right) \ket{\psi \left( t \right)}.
\label{eq:Wave Function}
\end{equation}
Its norm can be expressed with the jump probability $\delta p$ as
\begin{equation}
\label{eq:Probability Time Step delta t}
\langle \psi' \left( t+\delta t \right) \ket{\psi' \left( t+\delta t \right)}=1-\delta p,
\end{equation}
with
\begin{multline}
\delta p = \delta t \frac{\mi}{\hbar} \bra{\psi \left( t \right)} \hat{H}_{\textrm{nH}} - \hat{H}^{\dagger}_{\textrm{nH}} \ket{\psi \left( t \right)} \\
= \delta t\,\Gamma\,\bra{\psi \left( t \right)} \hat{C}^+ \hat{C}^- \ket{\psi \left( t \right)}.
\label{eq:Probability Quantum Jump}
\end{multline}
Since this stepwise MCWF method is first order in the sense that it allows for at most one quantum jump per time step, the stepsize $\delta t$ has to be small enough that the condition $\delta p \ll 1$ is fulfilled, and hence the probability of two jumps occurring in the same time step ($\sim(\delta p)^2$) is negligible.

At time $t+\delta t$, we switch to the state vector
\begin{equation}
\ket{\psi\left( t+ \delta t \right)}_\text{jump} \propto \hat{C}^- \ket{\psi'\left( t + \delta t\right)}
\label{eq:Evolution Jump Infinitesimal}
\end{equation}
with probability $\delta p$, which means that a quantum jump has occured. The state becomes
\begin{equation}
\ket{\psi\left( t+ \delta t \right)}_\text{no-jump} = \frac{1}{ \sqrt{1-\delta p}} \ket{\psi' \left( t+ \delta t\right)}
\end{equation}
with the complementer probability $1-\delta p$.

\subsection{Emission from a coherent superposition}
\label{sec:Emission From Coherent Superposition}
The workings of the method is exemplified with a two-level system with the Hamiltonian $\hat{H} = \hbar \omega_0 \hat{\sigma}^+ \hat{\sigma}^-$, where the general operators in Eq.~\eqref{eq:Relaxation Operator} are replaced by $\hat{C}^- = \hat{\sigma}^-$ and $\hat{C}^+ = \hat{\sigma}^+$. This is just the Dicke superradiance model introduced in Sec. \ref{sec:Emission rate equations with Dicke states} for $N=1$. The most general pure initial state reads
\begin{equation}
\ket{\psi \left( 0 \right)} = \alpha_0 \ket{g} + \beta_0 \ket{e}.
\label{eq:Initial State Quantum Jump}
\end{equation}

If we assume no quantum jump between $0$ and $t$, the normalized wave function at the general time $t$ can be written as
\begin{subequations}
\label{eq:Evolution No Jump}
\begin{equation}
\ket{\psi \left( t \right)} = \alpha \left( t \right) \ket{g} + \beta \left( t \right) \me^{-\mi \omega_0 t} \ket{e},
\end{equation}
where $\alpha \left( t \right)$ and $\beta \left( t \right)$ can be computed as solutions of the non-unitary evolution with Hamiltonian~\eqref{eq:Total Hamiltonian} and read \cite{Molmer:1993}
\begin{align}
\alpha \left( t \right) &= \frac{\alpha_0}{\sqrt{\left| \alpha_0 \right|^2 + \left| \beta_0 \right|^2 \me^{-\Gamma t}}},\\
\beta \left( t \right) &= \frac{\beta_0 \me^{-\frac{\Gamma t}{2}}}{\sqrt{\left| \alpha_0 \right|^2 + \left| \beta_0 \right|^2 \me^{-\Gamma t}}}.
\end{align}
\end{subequations}
With the help of these equations, one can derive the decaying probability of having no quantum jump between $0$ and $t$ as
\begin{equation}
p \left\{\text{no jump before } t \right\} = |\alpha_0|^2 + |\beta_0|^2 \me^{-\Gamma t}.
\label{eq:Probability No Jump}
\end{equation} 
For the derivation cf. \footnote{%
Eq. \eqref{eq:Probability No Jump} can be derived from the formula for the probability that an event with time-dependent occurence rate $r(\tau)$ does not occur before time $t$:%
\[p\{\text{occurence later than }t\}=\exp\left(-\int_0^td\tau\;r(\tau)\right).\]
This result can be derived in a very similar way as the exponential distribution for constant occurence rate. Eq. \eqref{eq:Probability No Jump} can be obtained by substituting \eqref{eq:Jump rate from superposition} into this equation.} Taking the limit $t \rightarrow \infty$, it becomes clear that with a probability $|\alpha_0|$ no jump will ever occur during the evolution, the system goes to the state $\ket{g}$ continuously. On the other hand, the probability density of a jump occurring reads 
\begin{equation}
p (t) = \Gamma\,|\beta_0|^2\,e^{-\Gamma t},
\label{eq:Proba Density Single Emitter}
\end{equation}
which is nothing else than the exponential waiting-time distribution for a process with constant occurrence rate $\Gamma$, except that it is normalized to $|\beta_0|^2$ instead of 1. This is despite the fact that the jump rate is not constant, but from Eq. \eqref{eq:Probability Quantum Jump} decreases according to 
\begin{multline}
r(t)=\frac{\delta p}{\delta t} = \Gamma\,\bra{\psi \left( t \right)} \hat{\sigma}^+ \hat{\sigma}^- \ket{\psi \left( t \right)}\\=
\frac{\Gamma\,|\beta_0|^2\,\me^{-\Gamma t}}{\left| \alpha_0 \right|^2 + \left| \beta_0 \right|^2 \me^{-\Gamma t}}.
\label{eq:Jump rate from superposition}
\end{multline}

In summary, starting from the initial condition \eqref{eq:Initial State Quantum Jump}, in an $|\alpha_0|$ fraction of the cases, no jump will ever be observed, whereas in the remaining $|\beta_0|$ fraction, a jump will be observed with probability density $\Gamma\,e^{-\Gamma t}$. In the $t\to\infty$ limit, the state is $\ket g$ in both cases. The exact same behavior is observed if the initial state is the completely mixed state
\begin{equation}
\rho(0)=|\alpha_0|\,\ket g \bra g + |\beta_0|\,\ket e \bra e.
\end{equation}
A mixed initial state can be translated to the MCWF method in such a way that the trajectories of the statistical ensemble are started not from the same state as above, but an $|\alpha_0|$ fraction of the trajectories is started from the state $\ket g$ (whence no jump will ever be observed), whereas the remaining $|\beta_0|$ fraction from the state $\ket e$ (whence a jump will be observed with probability density $\Gamma\,e^{-\Gamma t}$). This is a central result of the present paper, that was hereby proven for the case of a single emitter in the language of MCWF trajectories that with time-resolved observation of the collective emission burst, it is impossible to distinguish between coherent superposition and mixture in this scenario. A very similar calculation is possible for 2 emitters, as we demonstrate in Appendix \ref{app:TwoEmitters}.

\subsection{Implementation}

For the MCWF simulations in this work we use C++QED: a framework for simulating open quantum systems, where simple quantum mechanical operators can be pieced together to form complex systems. This software leverages C++ to yield high-performance executables meant for large-scale data collection, often in supercomputing environments. There are different options to simulate time evolutions of these systems, namely single MCWF trajectories, ensembles of many trajectories, and master equations. The basic idea and ways of implementation are discussed in Ref.~\cite{Vukics:2007}. An updated version, C++QEDv2 \cite{Vukics:2012, Sandner:2014} is currently maintaned and available as an open-source package.

Based on the general concept of the MCWF method, there are several tools implemented in C++QED to improve the quality of the simulation. Most importantly, it is possible to allow for an adaptive timestep, that is essential for production-scale numerics. The timestep is varied by an adaptive ordinary differential equation solver in order to guarantee a preset precision level for the continuous non-unitary evolution defined by the Hamiltonian \eqref{eq:Total Hamiltonian}. Meanwhile, a superimposed mechanism monitors that the jump probability per timestep \eqref{eq:Probability Quantum Jump} remains very small, by limiting the stepsize $\delta t$ from above. Namely, a parameter $\Delta p\ll1$ is introduced, and it is made sure that
\begin{equation}
 \delta p < \Delta p\quad\text{at all times.}
\end{equation}
This makes sure that the probability of two jumps occurring in the same timestep (which possibility is not included in the 1st order MCWF method that we use here) is negligible, namely, less than $(\Delta p)^2$.

\section{Comparison of Methods}
\label{sec:Comparison of the methods}
Whereas the rate-equation model \eqref{eq:System of Rate Equations} operates on the diagonal of the density matrix ($N+1$ entries for $N$ atoms), so that it cannot describe quantum coherence, the MCWF method works with statistical ensembles of full state vectors with $N+1$ entries, so that the latter method is much more demanding numerically.

We use these two methods to study the intensity $I \left( t \right)$ with various initial conditions and test them for the conditions of superradiance. Beside the photon emission from an initially completely excited Dicke state \eqref{eq:Initial State}, we look at the radiation from a statistical mixture of half of the atoms initially excited and the other atoms deexcited (system of rate equations). This is contrasted to a symmetrical superposition of an entirely excited state and a deexcited state (MCWF method) and used to check the accuracy of the two methods.

To see how the MCWF method works in this case, let us consider an initial state of equal weight between the maximally excited Dicke state \eqref{eq:Initial State} $\ket{J,J} = \ket{e,e,e,\dots,e}$ and the completely deexcited state $\ket{J,-J} = \ket{g,g,g,\dots,g}$
\begin{equation}
\ket{\psi \left( 0 \right)} = \frac{1}{\sqrt{2}} \left( \ket{J,J} + \ket{J,-J} \right).
\label{eq:Initial State Coherent Superposition}
\end{equation}
We make the following substitutions $\hat{C}^- = \hat{J}^-$ and $\hat{C}^+ = \hat{J}^+$.

The evolution of an initial state $\ket{\psi \left( 0 \right)}$ is governed by both continuous decay due to the non-Hermitian Hamiltonian \eqref{eq:Total Hamiltonian} and the jump part $\hat{J}^- \hat{\rho} \hat{J}^+$, which is responsible for the emission of photons, as explained in Sec.~\ref{sec:Monte Carlo wave functions}. The Dicke states $\ket{J,M}$ \eqref{eq:Dicke State} are eigenstates of the non-Hermitian Hamiltonian \eqref{eq:Total Hamiltonian}. Since the state $\ket{\psi \left( t \right)}$ is continuously renormalized during the time evolution \cite{Kornyik:2019}, the non-Hermitian Hamiltonian does not affect Dicke states at all in the course of the dynamics. As a consequence, Dicke states do not undergo continuous decay, but can decay through jumps only.

Since the coherent superposition of Dicke states \eqref{eq:Initial State Coherent Superposition} is not an eigenstate of the non-Hermitian Hamiltonian \eqref{eq:Total Hamiltonian}, it decays both in a continuous decay and by jumps. Until the first jump occurs, the coherent superposition state \eqref{eq:Initial State Coherent Superposition} evolves according to the non-Hermitian Hamiltonian \eqref{eq:Total Hamiltonian} as (normalization included)
\begin{equation}
\ket{\psi \left( t \right)} = \frac{1}{\sqrt{1+\me^{-N \Gamma t}}} \left( \ket{J,-J} + \me^{-\frac{1}{2} N \Gamma t} \ket{J,J} \right).
\end{equation}
This behavior is equivalent to Eq.~\eqref{eq:Evolution No Jump} except for the increased damping rate. One can see that the excited component is damped gradually due to the non-Hermitian evolution. The jump rate reads
\begin{equation}
r \left( t \right) = N\, \Gamma \frac{\me^{-N \Gamma t}}{1 + \me^{-N \Gamma t }},
\label{eq:Jump Rate First Jump}
\end{equation}
which corresponds to Eq.~\eqref{eq:Probability Quantum Jump}. Just like in the single-emitter case above, the occurrence of a jump becomes the less likely, the longer one waits for the jump to happen. Its probability even converges to $0$ for $t \to \infty$.

The first jump, given by the application of the operator $\hat{J}^-$, annihilates the deexcited component of the initial state \eqref{eq:Initial State Coherent Superposition} $\ket{J,-J}$ and reduces the number of excitations in $\ket{J,J}$ by one, bringing the system to the Dicke state $\ket{J,J-1}$, cf. Eq.~\eqref{eq:Evolution Jump Infinitesimal}
\begin{equation}
\ket{\psi ( t )}_\text{jump} \propto \hat{J}^- \ket{\psi \left( t \right)} \propto \ket{J,J-1}.
\end{equation}
From this point on, the system behaves as if the initial state was a Dicke state and can lose its excitations only in a series of jumps. Thus it becomes clear that there are two types of trajectories, each occurring with probability $1/2$: trajectories of continuous decay and trajectories of (a series of) jumps. It is also apparent that the occurrence of the first jump determines whether the trajectory is of the continuous-decay or the series-of-jumps type.

Therefore the question arises whether there is a fundamental difference between the time evolution of a quantum state starting with the coherent superposition \eqref{eq:Initial State Coherent Superposition} and the statistical mixture starting from the completely mixed state
\begin{equation}
\hat{\rho} \left( t \right) =\frac12\left( \ket{J,J}\bra{J,J} + \ket{J,-J}\bra{J,-J} \right).
\label{eq:Completely Mixed Initial Condition}
\end{equation}
The emission dynamics of such a state can be treated with the classical rate equations \eqref{eq:System of Rate Equations}, and the total number of emitted photons is $N/2$, which is identical to the expectation value for the number of jumps on a quantum trajectory started from the superposition \eqref{eq:Initial State Coherent Superposition}.

Since a coherent superposition exhibits nonvanishing off-diagonal density-matrix elements in contrast to the completely mixed state, it is \emph{a priori} not clear whether the dynamics may be different for the two cases. However, given our derivation above, such differences can only stem from the first quantum jump. In the following, we answer this question by looking at the waiting-time distribution of the first jump. Based on the jump rate for the first jump $r \left( t \right)$ \eqref{eq:Jump Rate First Jump}, the probability density is obtained in the same way as Eq. \eqref{eq:Proba Density Single Emitter} to read
\begin{equation}
p\left( t \right) = \frac{N}{2} \Gamma \exp \left( - N \Gamma t \right).
\label{eq:Waiting Time Density}
\end{equation}
Its norm being $1/2$ means that there is no emitted photon in half of the trajectories, which is the same as if the system was started from the pure state $\ket{J,-J}$, that is, the second term in the mixture \eqref{eq:Completely Mixed Initial Condition}. The other half of the trajectories yield the first emitted photon with (conditional) waiting-time density $2 p ( t ) = N \Gamma \exp \left( -N \Gamma t \right)$, which is the same as if the system was started from the pure state $\ket{J,J}$, that is, the first term in the mixture \eqref{eq:Completely Mixed Initial Condition}. As discussed above, after the first photon is emitted, the trajectory behaves the same as if it was started from the pure state $\ket{J,J-1}$. Hence, the conclusion here is the same as in Sec. \ref{sec:Emission From Coherent Superposition} for the single-emitter case: by time-resolved photon counting, it is not possible to distinguish between the initial conditions \eqref{eq:Initial State Coherent Superposition} and \eqref{eq:Completely Mixed Initial Condition}.

Let us see how the radiation intensity can be calculated from the MCWF method. Since each quantum jump is equivalent to a photon leaving the atomic ensemble, the intensity during a time interval $\delta t$ at time $t$ can be defined as
\begin{equation}
\label{eq:IntensityMCWF}
I_\text{MCWF}(t)=\frac{\text{\small number of quantum jumps between $t$ and $t+\delta t$}}{\delta t}.
\end{equation}
Hence, the time-resolved intensity from the MCWF method is nothing else than a temporal histogram of quantum jumps, and the better the time resolution, the more trajectories we need for acceptable accuracy of the histogram.

\begin{figure}
	\centering
    \includegraphics[width=\columnwidth]{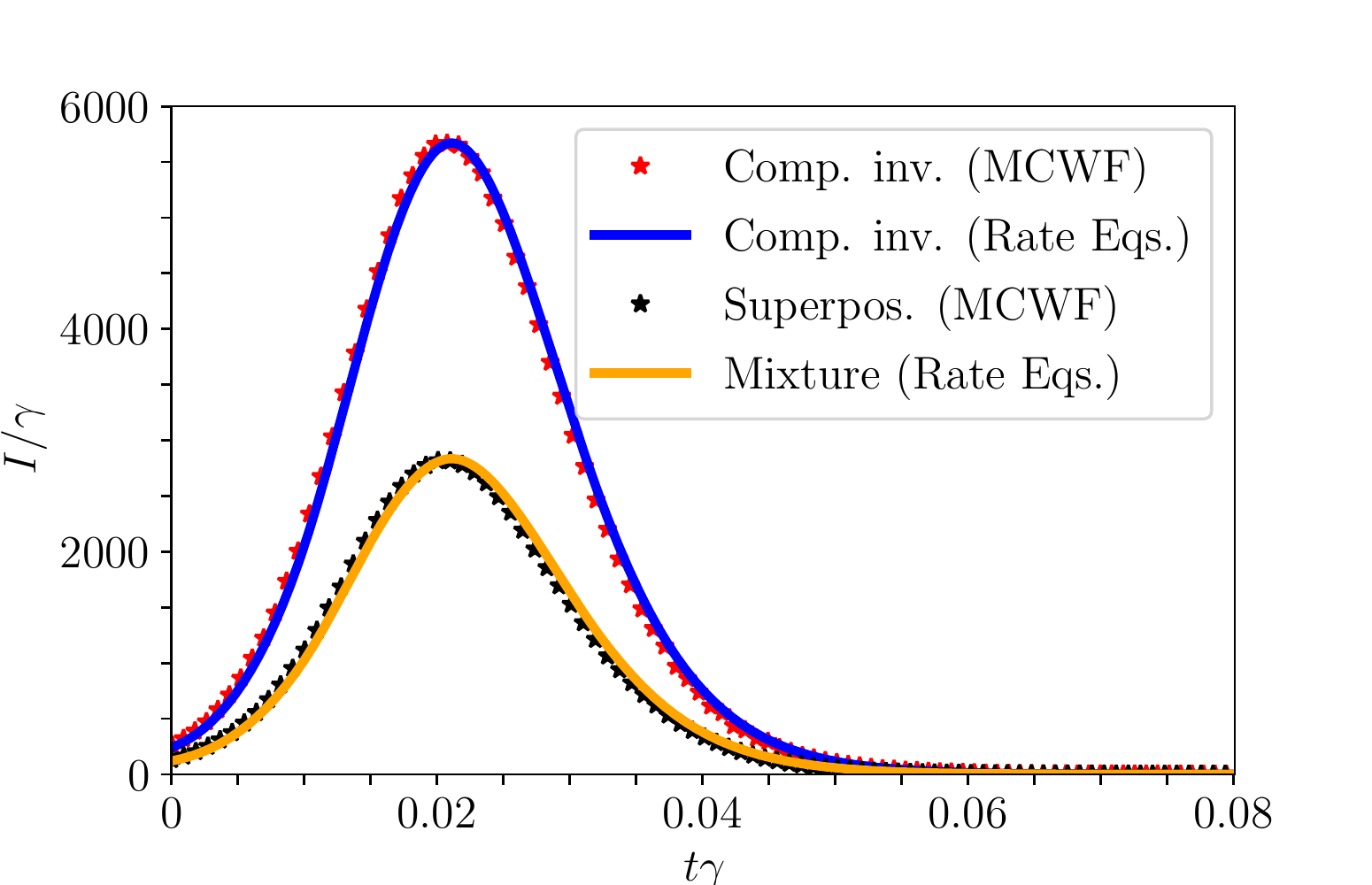}\par
    \bigskip
    {\renewcommand{\arraystretch}{1.5}
\begin{tabular}{|p{2.1cm}|p{1.45cm}|p{1.45cm}|p{1.45cm}|p{1.45cm}|}
\hline
Method & Peak height & Peak time & Number of jumps & Int. photon emission\\
\hline
\hline
Rate Equations & $2834\,\gamma$ & $0.021/\gamma$ & - & $59.9$\\
\hline
MCWF & $2821\,\gamma$ & $0.020/\gamma$ & $59.6$ & $59.5$\\
\hline
\end{tabular}}
	\caption{Comparison between the classical emission rate equations method with an initial density matrix of $\hat{\rho} \left( 0 \right) = \ket{J,J} \bra{J,J}$ and the MCWF approach with the initial state $\ket{\psi \left( 0 \right)} = \ket{J,J}$ \eqref{eq:Initial State} for $N=120$ atoms. Additionally, we compare the rate-equation method from the initial density matrix of $\hat{\rho} \left( 0 \right) = \frac{1}{2} \left( \ket{J,J} \bra{J,J} + \ket{J,-J} \bra{J,-J} \right)$ with the MCWF method from the initial state $\ket{\psi \left( 0 \right)} = \frac{1}{\sqrt{2}} \left( \ket{J,J} + \ket{J,-J} \right)$ \eqref{eq:Initial State Coherent Superposition}. For the MCWF method $\Delta p=10^{-3}$, and the intensity is calculated for $100$ temporal bins over $10^4$ trajectories. Here, as in all the following figures, $\gamma=\Gamma/2$ is the scaling factor for both time and intensity. The table shows the value of the maximum of the intensity and its position in time in the case of the superposition/mixture initial state. The total number of jumps in the MCWF approach corresponds to the integrated photon emission and deviates slightly from the value of $60$ due to numerical inaccuracy.}
	\label{fig:Data}
\end{figure}

Fig.~\ref{fig:Data} exhibit numerical results of the rate-equation and MCWF models. The intensity is computed from Eq.~\eqref{eq:Emission Intensity} for the rate-equation model, and \eqref{eq:IntensityMCWF} for the MCWF method.
and show the numerically obtained intensity $I(t)$. As was proven theoretically, the emitted intensity follows the same temporal behavior for the initial conditions of equal superposition of the highest and lowest Dicke state (MCWF method) and equal mixture thereof (rate equations). Due to the error stemming from the finite $\Delta p$ in the MCWF method, the peak is slightly shifted and is smaller than in the rate-equation method.

\section{Superradiance for various initial states}
\label{sec:Superradiance for several initial states}
In a next step we compare different initial conditions, namely the semi-inverted Dicke state and the superposition of a completely inverted state and the ground state (\ref{sec:Comparison between semi-inverted Dicke state and Dicke state mix}). In addition, we investigate the dynamics starting from a superposition of the completely inverted state and the ground state with unequal weights (\ref{sec:Superposition state with unequal weight}). Finally, we split the atomic ensemble of $N$ atoms into two chunks. The collective atomic states in each chunk are Dicke states and follow the laws of superradiance. Since there is no coupling between the chunks, the entire system has a much bigger Hilbert space. We want to check in Sec.~\ref{sec:Collective emission from two chunks} how the superradiant emission burst behaves in this case and how it depends on the partitioning.

\subsection{Comparison between semi-inverted Dicke state and Dicke state mix}
\label{sec:Comparison between semi-inverted Dicke state and Dicke state mix}
\begin{figure}
	\centering
		\includegraphics[width=\columnwidth]{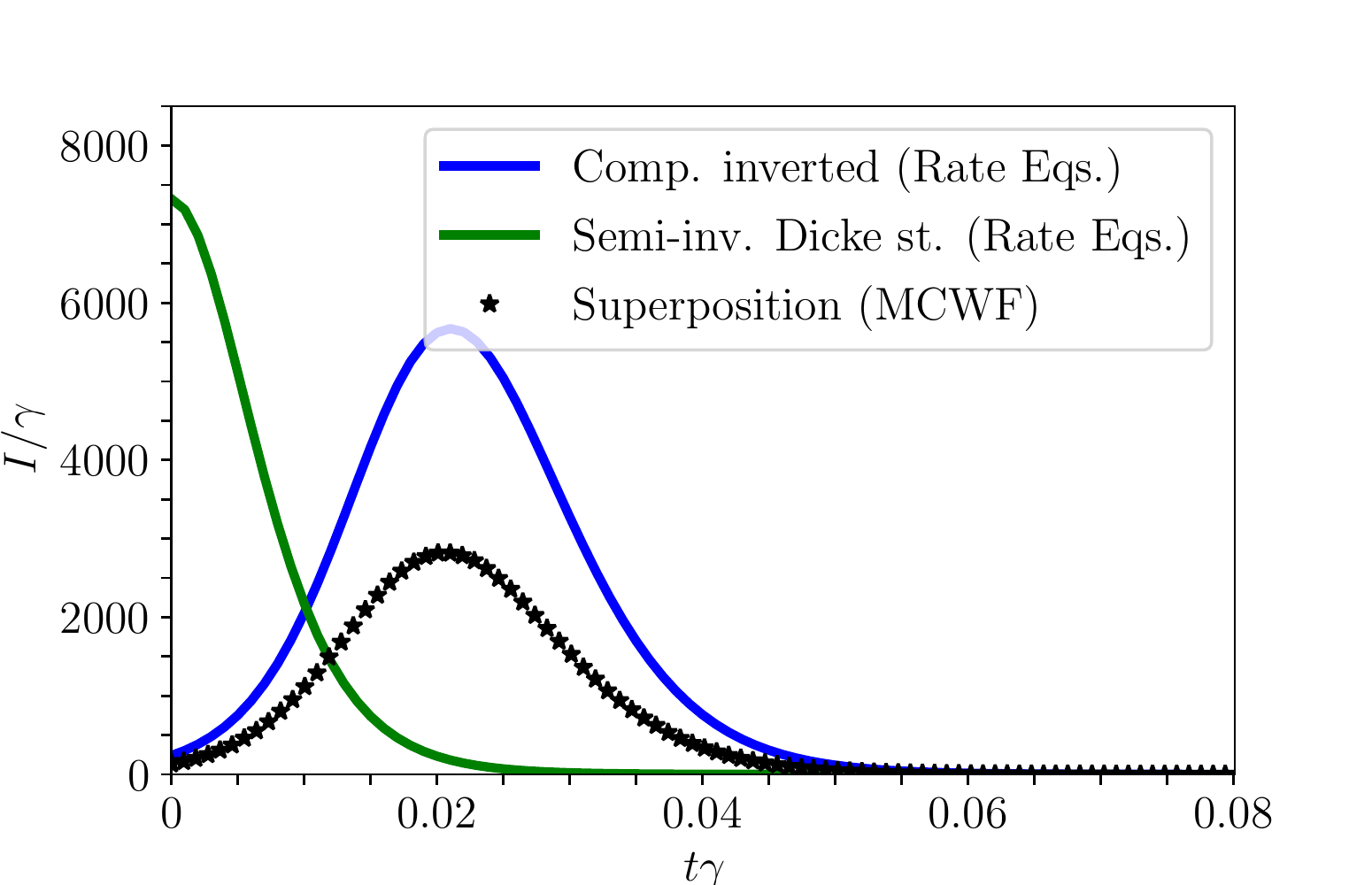}
	\caption{Collective emission for two different initial states: superposition of the completely inverted state and the ground state, and semi-inverted Dicke state. The temporal intensity curve for emission from the completely inverted state is plotted for reference.}
	\label{fig:InitialState}
\end{figure}

The initial state $\ket{\psi \left( 0 \right)} = \frac{1}{\sqrt{2}} \left( \ket{J,J} + \ket{J,-J} \right)$ and the semi-inverted Dicke state $\ket{\psi \left( 0 \right)} = \ket{J, M=0}$ possess the same number of excitations, but initiate a completely different dynamic behavior. Fig.~\ref{fig:InitialState} shows a comparison of dynamics. Whereas the first scenario leads to a peak of reduced height, the peak height of the semi-inverted Dicke state exceeds even the one of the completely inverted Dicke state $\ket{\psi \left( 0 \right)} = \ket{J,J}$. In case of a total number of $120$ atoms with $60$ excitations, the integrated intensity reproduces this latter number in either case. We conclude that the emission is faster for the semi-inverted Dicke state, which feature can be attributed to the higher symmetry of the state.


\subsection{Superposition state with unequal weight}
\label{sec:Superposition state with unequal weight}
Let us continue by considering the superposition with arbitrary (real) weight $c$:
\begin{equation}
\ket{\psi \left( 0 \right)} = \frac{1}{\sqrt{1+c^2}} \left( \ket{J,J} + c \ket{J,-J} \right).
\label{eq:Initial State Coherent Superposition c}
\end{equation}
We want to check if the collective emission starting from this initial state still fulfills the characteristics of superradiance. To this end, we choose various values of $c$ and plot the values of the peak height and the peak width as a function of the number of atoms $N$. As mentioned in Sec.~\ref{sec:Emission rate equations with Dicke states}, the peak height is supposed to scale with $N^2$ and the peak width with $1/N$.

\begin{figure}
	\centering
	\begin{tabular}{r l}
		(a) & \includegraphics[width=.95\columnwidth]{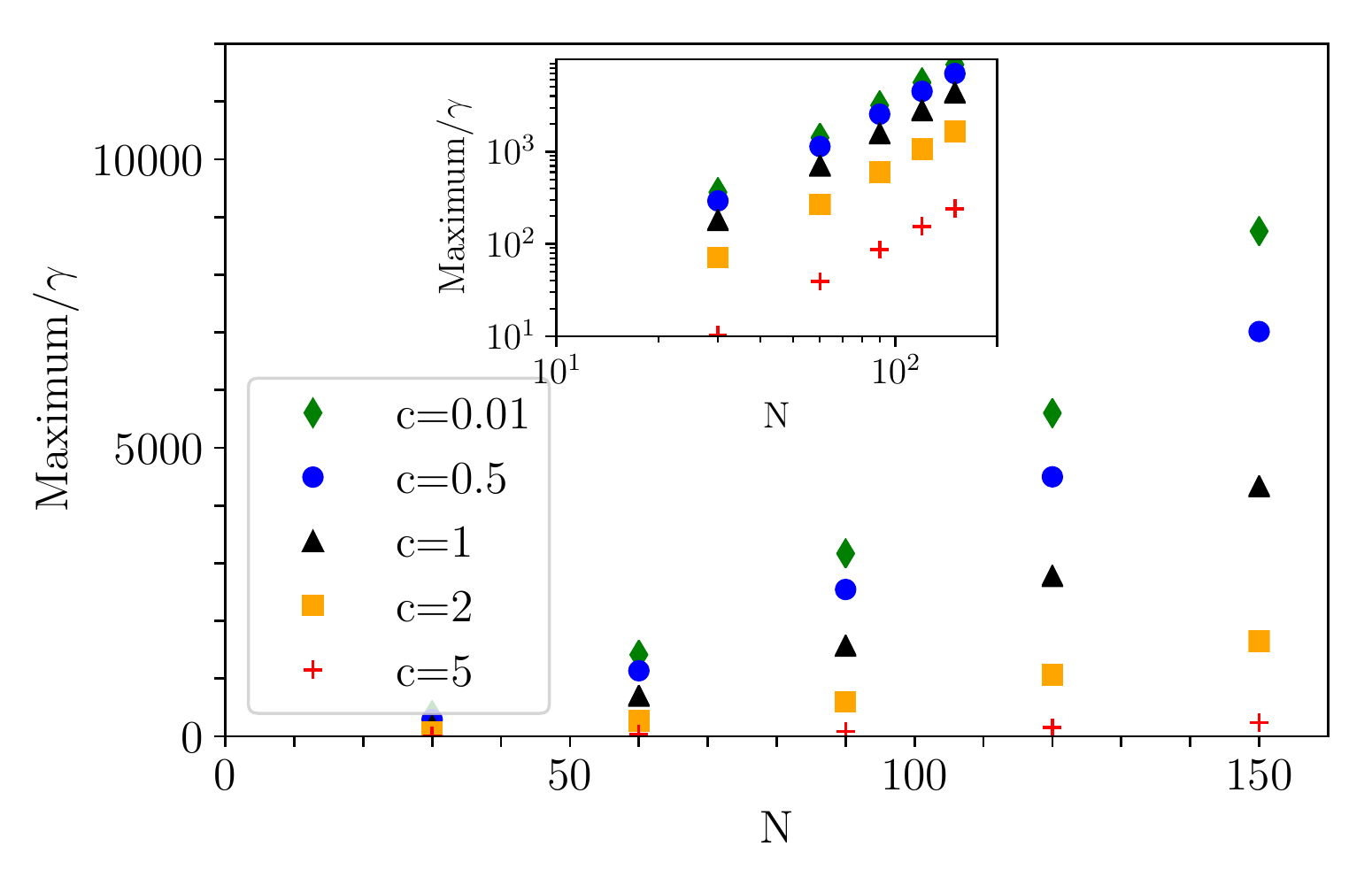}\\
		(b) & \includegraphics[width=.95\columnwidth]{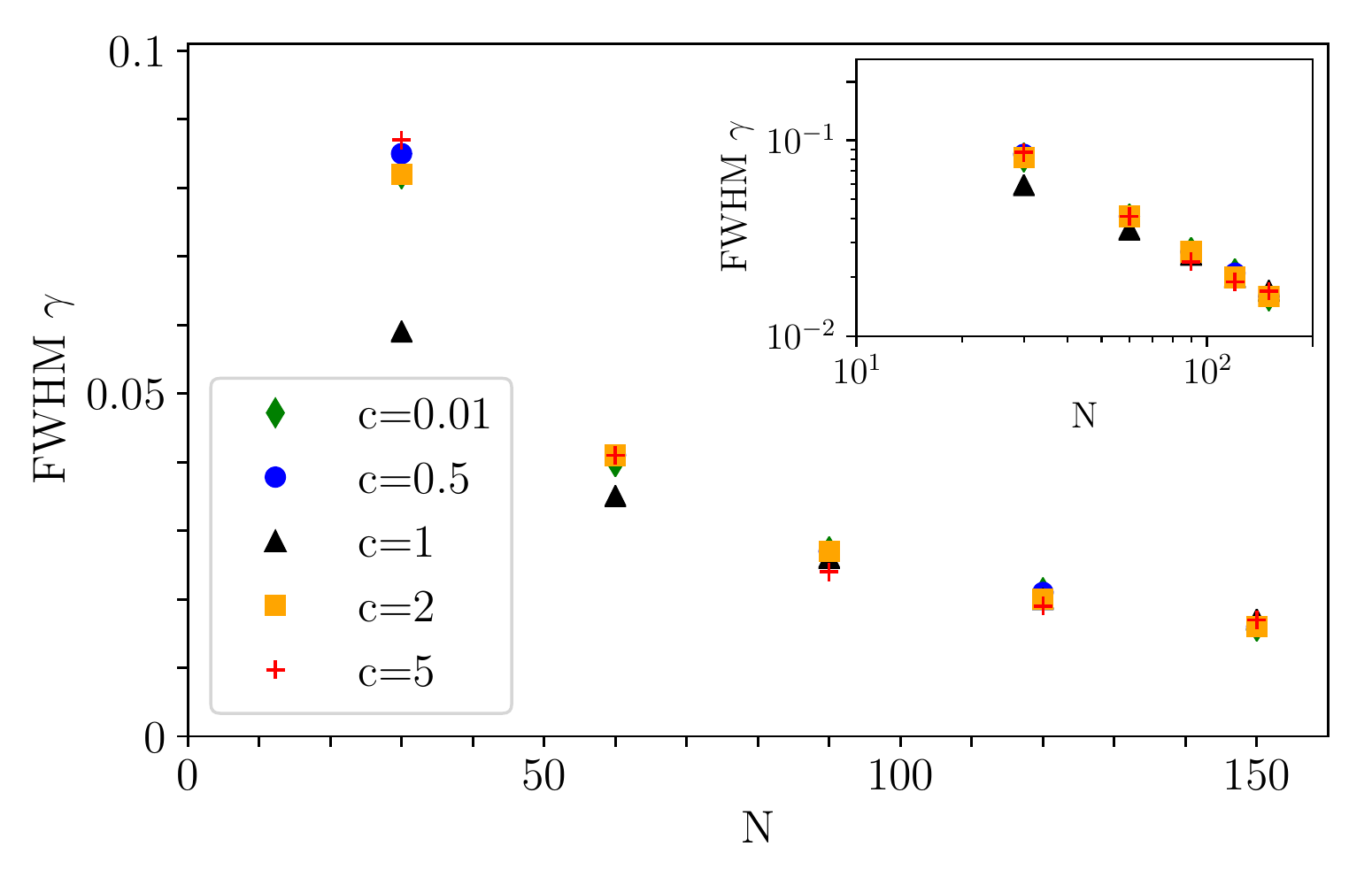}
    \end{tabular}
	\caption{(a) Peak height and (b) peak width (FWHM) of the intensity as a function of the number of atoms $N$ for several initial states parametrized by the value $c$ in Eq.~\eqref{eq:Initial State Coherent Superposition c}. The MCWF approach is used with a total number of $100$ bins over $10^4$ trajectories, and $\Delta p$ is set to $10^{-2}$. The parameters controlling the MCWF method remain the same in all the subsequent figures.}
	\label{fig:Maximumc}
\end{figure}

Figure ~\ref{fig:Maximumc} shows the peak height and the peak width (FWHM) as a function of the number of atoms $N$ for the values of $c$ in Eq.~\eqref{eq:Initial State Coherent Superposition c} with $c=0.01, \: 0.5, \: 1, \: 2$ and $5$. The insets shows logarithmic plots, and the exponents obtained from linear fits are displayed in Tab.~\ref{tab:Comparison of Initial States}.

\begin{table}
{\renewcommand{\arraystretch}{1.5}
\begin{tabular}{|p{1.7cm}|p{3cm}|p{3cm}|}
\hline
$c$ in Eq.~\eqref{eq:Initial State Coherent Superposition c} & Exp. of peak height & Exp. of FWHM\\
\hline
\hline
$0.01$ & $1.990$ & $-1.009$\\
\hline
$0.5$ & $1.985$ & $-1.034$\\
\hline
$1$ & $1.989$ & $-0.947$\\
\hline
$2$ & $1.974$ & $-1.012$\\
\hline
$5$ & $1.983$ & $-1.096$\\
\hline
\end{tabular}}
\caption{Fit parameters for the exponents of the peak height and peak width (FWHM) for several initial states given by the value $c$ in Eq.~\eqref{eq:Initial State Coherent Superposition c} as a function of the number of atoms $N$. We use atom numbers of $N=30, 60, 90, 120$ and $150$. The number of bins is set to $100$ and $\Delta p$ is $10^{-2}$.}
\label{tab:Comparison of Initial States}
\end{table}

The results of the fit parameters in Tab.~\ref{tab:Comparison of Initial States} are perfectly compatible with the conditions of superradiance. We conclude that collective emission processes from all initial states \eqref{eq:Initial State Coherent Superposition c} lead to superradiant emission features.

\subsection{Collective emission from two chunks}
\label{sec:Collective emission from two chunks}
In this section we study two alternative ways of increasing the number of emitting atoms. We introduce chunks of atoms separated by more than a radiation wavelength so that photons coming from different chunks are distinguishable, whereas within a single chunk, it cannot be distinguished which particle an emitted photon originates from. This means that the atoms within a single chunk form Dicke states, but there is no symmetrization between the states of two chunks. The result will be that only atoms within a single chunk will emit cooperatively. The total number of atoms is given by the product of the number of chunks $N_{\textrm{Ch}}$ and the number of particles per chunk $N_{\textrm{PPCh}}$: $N = N_{\textrm{Ch}} \sprod N_{\textrm{PPCh}}$.

\begin{figure}
	\centering
		\includegraphics[width=\columnwidth]{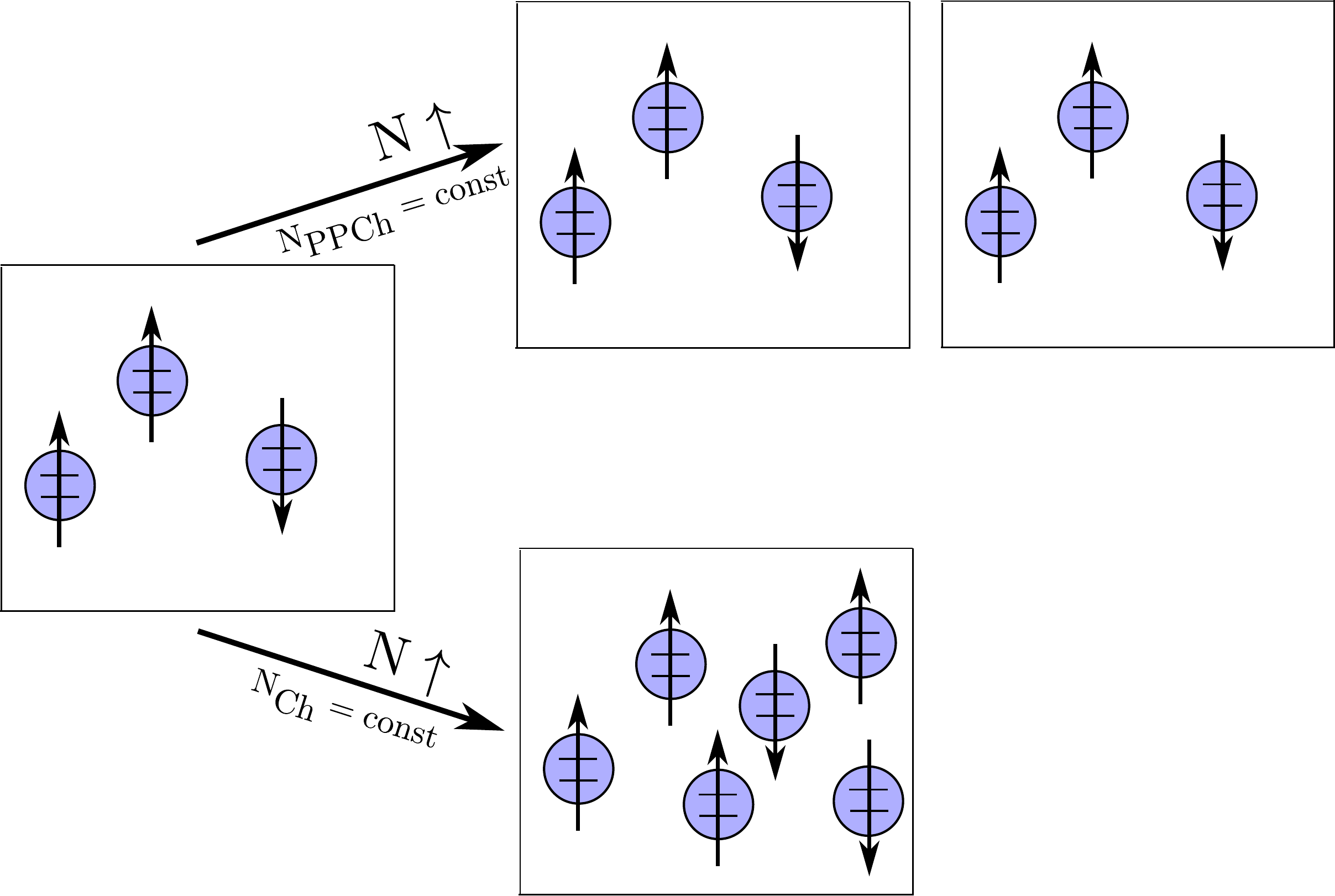}
	\caption{There are two different ways of increasing the number of emitting atoms. The number of chunks $N_{\textrm{Ch}}$ can be increased and the number of atoms per chunk $N_{\textrm{PPCh}}$ can be kept fixed. This situation is shown in the upper half. The other way is to keep the number of chunks $N_{\textrm{Ch}}$ constant while increasing the number of atoms per chunk $N_{\textrm{PPCh}}$, which is depicted in the lower half of the figure.}
	\label{fig:Sketch_Chunks}
\end{figure}

Firstly, the particle number can be changed by increasing the number of chunks while keeping the number of particles per chunk constant, corresponding to constant particle density. This situation is depicted in the upper half of Fig.~\ref{fig:Sketch_Chunks}. The the peak intensity $I_{\textrm{p}}$ is proportional to the number of chunks $N_{\textrm{Ch}}$, but proportional to the number of particles per chunk squared $N^2_{\textrm{PPCh}}$. Thus, intensity scales linearly with the total number of atoms $N$:
\begin{equation}
I_{\textrm{p}} \propto N_{\textrm{Ch}}\cdot N^2_{\textrm{PPCh}}
= \frac{N}{N_{\textrm{PPCh}}} N^2_{\textrm{PPCh}} \propto N,
\end{equation}
meaning that in this situation we have no superradiance.

Alternatively, the number of chunks $N_{\textrm{Ch}}$ is kept constant, cf. lower half of Fig.~\ref{fig:Sketch_Chunks}, corresponding to a constant volume for the particles. In this scenario the emitted photon cannot be assigned to any single atom within a single chunk. The peak intensity $I_{\textrm{p}}$ behaves as
\begin{equation}
I_{\textrm{p}} \propto N_{\textrm{Ch}}\cdot N^2_{\textrm{PPCh}}
= N_{\textrm{Ch}} \frac{N^2}{N^2_{\textrm{Ch}}} \propto N^2
\label{eq:Intensity quadratic}
\end{equation}
showing the expected superradiant behavior.

Our initial state is a product state of coherent superpositions of two Dicke states in each chunk
\begin{equation}
\ket{\psi \left( 0 \right)} = \frac{1}{2^{\frac{N_\text{Ch}}2}}\bigotimes_i \big( \ket{J,J}_i + \ket{J,-J}_i\big)
\label{eq:Initial State Two Chunks}
\end{equation}
where $i$ indexes the chunks.

\begin{figure}
	\centering
	\begin{tabular}{r l}
	(a)&\includegraphics[width=.95\columnwidth]{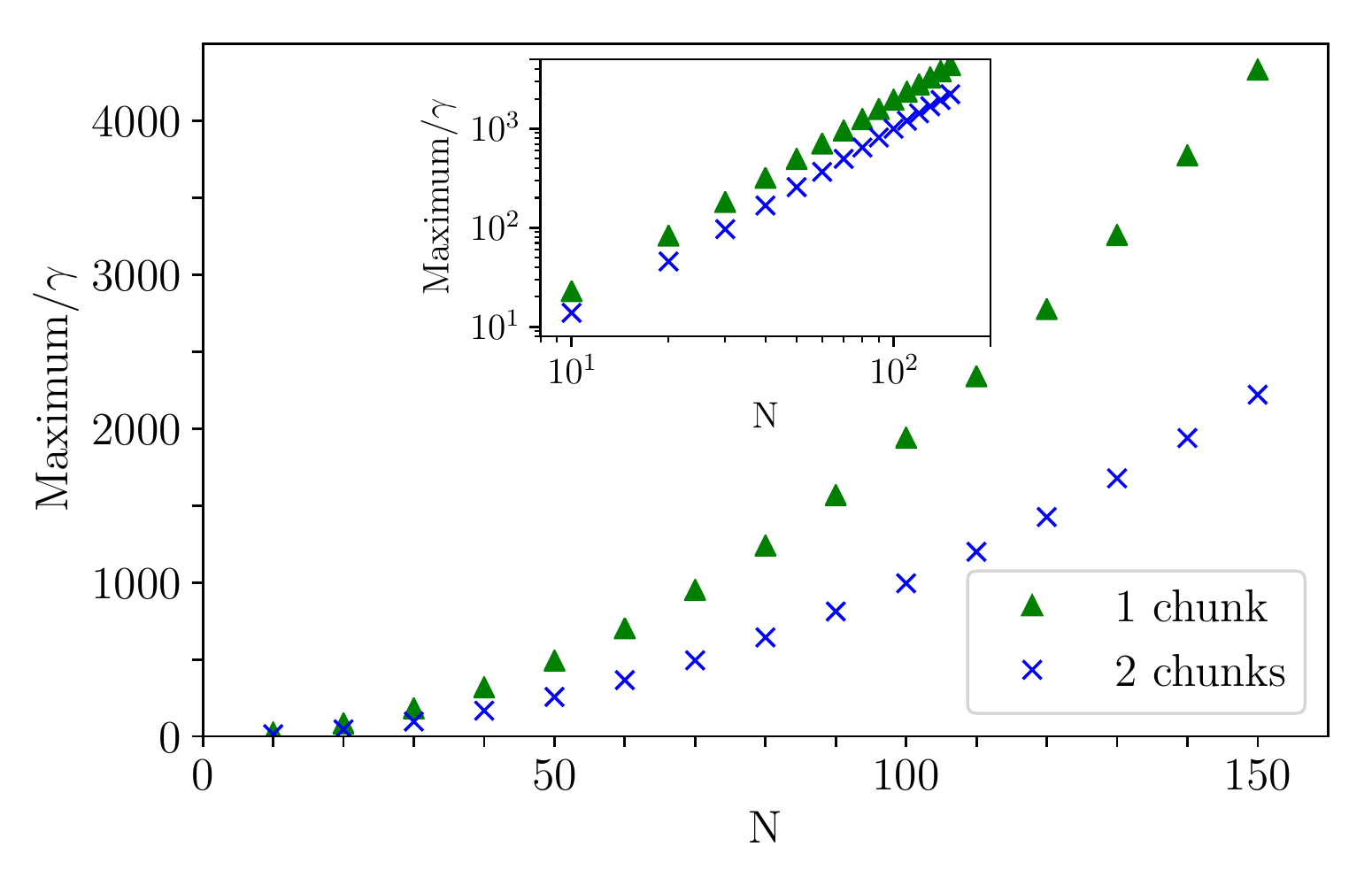}\\
	(b)&\includegraphics[width=.95\columnwidth]{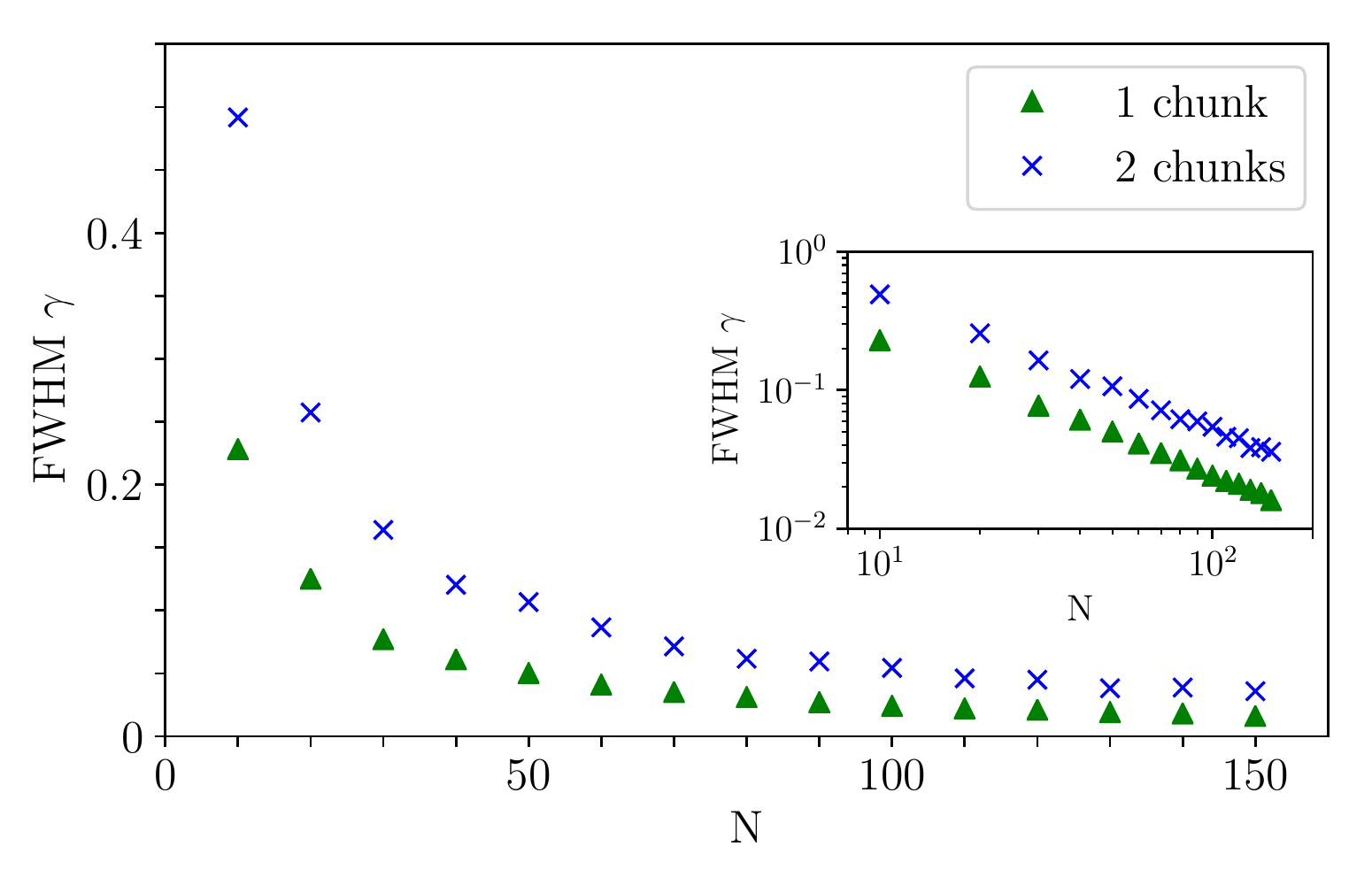}
	\end{tabular}
	\caption{(a) Peak height and (b) peak width of the intensity $I(t)$ from the MCWF approach as a function of the total number of atoms in the two chunks $N$ for an initial state given in Eq.~\eqref{eq:Initial State Two Chunks} for one and two chunks.}
	\label{fig:MaximumCh}
\end{figure}

We concentrate on the second case in our analysis and keep the number of chunks $N_{\textrm{Ch}}$ constant. Figure \ref{fig:MaximumCh}(a) compares the peak height of the collective emission burst for the case of one and two chunks as a function of the total number of atoms $N$. The logarithmic plot in the inset of Fig.~\ref{fig:MaximumCh}(a) shows a linear curve confirming Eq.~\eqref{eq:Intensity quadratic}. Figure \ref{fig:MaximumCh}(b) shows the peak width (FWHM) of the intensity $I(t)$ \eqref{eq:Emission Intensity} for the case of one and two chunks as a function of the total number of atoms $N$ in the two chunks. The logarithmic plot in the inset again reveals the superradiant behavior.

\section{Summary}
We have shown that the MCWF approach is appropriate to investigate the superradiant decay of Dicke states. It has been proven analytically that the decay of an initial coherent superposition state of the maximally excited Dicke state $\ket{J, J}$ and the ground state $\ket{J, -J}$, $\ket{\psi \left( 0 \right)} = \frac{1}{\sqrt{2}} \left( \ket{J,J} + \ket{J,-J} \right)$, using the MCWF approach is in accord with the time evolution of a system of coupled classical rate equations starting from $\hat{\rho} \left( 0 \right) = \frac{1}{2} \left( \ket{J,J} \bra{J,J} + \ket{J,-J} \bra{J,-J} \right)$. This agreement was confirmed numerically, where the precision of the MCWF approach plays a central role. This agreement leads to the conclusion that with the time-resolved observation of the superradiant burst, it is not possible to distinguish between coherent superposition and mixture in this scenario. This result required the analysis of the time-dependent probabilities and conditional states in the quantum trajectory approach. Trajectories with less observed photons than the maximum number of atomic excitations in the superposition -- no-photon trajectories in the single-atom case, and 0- and 1-photon trajectories in the two-atom case -- play a central role here.

A comparison with the dynamics of the rate equation model starting from the semi-inverted Dicke state $\ket{J, 0}$ gives an insight into the complexity of superradiance. Here, the total number of excitations is the same as in the former scenario, however, the decay of the atoms is cooperative from the onset, which results in a very different temporal intensity curve.

In a next step, we studied the decay from initial states $\ket{\psi \left( 0 \right)} = \frac{1}{\sqrt{1+c^2}} \left( \ket{J,J} + c \ket{J,-J} \right)$ parametrized by the real parameter $c$, using the MCWF approach. We have found that the temporal emission intensity curve exhibits the features of superradiance given by the characteristic peak height and width for any value $c$.

Moreover, the emission dynamics in two separate chunks with initial state $\ket{\psi \left( 0 \right)} = \frac{1}{2} \left( \ket{J,J}_1 + \ket{J,-J}_1 \right) \otimes \left( \ket{J,J}_2 + \ket{J,-J}_2 \right)$ was investigated. There is no cooperation in the decay between the atoms from the different chunks, since these are separated by more than a wavelength. As a result, the characteristics of superradiance stemming from individual chunks is simply added up to the total emission. Thus the superradiant scaling behavior is governed not by the total number of atoms, but only by the number of atoms in each sub-wavelength chunk.

One can think of several reasons in an experiment studying the collective decay of emitters why superradiance might be diminished. If the laser used to initially excite the emitters manages to excite only a certain ratio of the atoms to the excited state, superradiance is still likely to be observed as we have proven. In contrast, in case of a large distance between emitters, when there are no collective ties between the contributors, only the sum of the emitted radiation can be measured. However, we have shown that the cooperative emission of only two atoms is theoretically sufficient for superradiance to occur.

\section{acknowledgments}
This work was supported by the German Research Foundation (DFG, Grants BU 1803/3-1 and GRK 2079/1) and the National Research, Development and Innovation Office of Hungary (NKFIH) within the Quantum Technology National Excellence Program (Project No. 2017-1.2.1-NKP-2017-00001) and by Grant No. K115624. A.~V.~acknowledges support from the J\'{a}nos Bolyai Research Scholarship of the Hungarian Academy of Sciences and valuable exchange with M. Kor\-nyik.


\appendix
\section{The most fundamental manifestation of superradiance}
\label{sec:DickePaper}

This is a modern reformulation of the argument given for neutrons in the introduction of Dicke’s original paper~\cite{Dicke:1954}.

\paragraph{Single emitter}
Let us consider a single emitter with two states $\ket{g}$ and $\ket{e}$ governed by the Hamiltonian $\hat{H} = \hat{H}^{(0)} + \hat{V}$ with the term $\hat{V}$ causing transition from $\ket{e}$ to $\ket{g}$ (in the Dicke model, this is the interaction with all the electromagnetic modes surrounding the emitter). The respective transition probability can be given as
\begin{equation}
p_{\textrm{one}} \propto \left| \bra{g} \hat{V}{}\ket{e} \right|^2.
\end{equation}

\paragraph{Two emitters in triplet state}
This is contrasted to the situation where two emitters $a$ and $b$, one in state $\ket{e}$ and the other in state $\ket{g}$, are in close vicinity, so that their state must be symmetrized. Note that the total number of excitations remains 1, as in the previous paragraph. The total Hamiltonian is here given by $\hat{H}{}_{\textrm{tot}} = \hat{H}^{(0)}_{a} + \hat{H}^{(0)}_{b} + \hat{V}{}_{a} + \hat{V}{}_{b}$. The state of the emitters reads $\ket{1,0} = \frac{1}{\sqrt{2}} \left( \ket{e, g} + \ket{g, e} \right)$, cf. Eq.~\eqref{eq:Dicke State}, the total transition probability of the entire system to the lowest state $\ket{g, g}$ is found
\begin{multline}
p_{\textrm{triplet}} \propto \left| \bra{g, g} \hat{V}{}_{a} + \hat{V}{}_{b} \ket{1,0} \right|^2 \\
= \left| \frac{1}{\sqrt{2}} \left( \bra{g, g} \hat{V}{}_{a} \ket{e, g} + \bra{g, g} \hat{V}{}_{b} \ket{g, e} \right) \right|^2 \\
= \frac{1}{2} \left| 2 \bra{g, g} \hat{V}{}_{a} \ket{e, g} \right|^2 = 2 p_{\text{single}},
\end{multline}
where the second equality holds due to symmetry between the two identical emitters. It is apparent that the presence of a second identical but deexcited emitter doubles the transition rate simply as a result of symmetrization (which introduces quantum coherence, and hence interference which is constructive in this case), even without any interaction between the two.

\paragraph{Singlet state} An equivalent calculation for the singlet state $\ket{0,0} = \frac{1}{\sqrt{2}} \left( \ket{e, g} - \ket{g, e} \right)$ leads to the transition probability of
\begin{equation}
p_{\textrm{singlet}}=0
\end{equation}
due to interference, which is destructive in this case.

\paragraph{The case of independent emitters} If, on the contrary, the particles are considered independent, which situation can be described by the state $\ket{e, g} = \frac{1}{\sqrt{2}} \left( \ket{1,0} + \ket{0,0} \right)$, we recover the transition probability
\begin{equation}
p_{\textrm{independent}}=p_{\textrm{one}}.
\end{equation}

\section{Emission from a coherent superposition state for 2 emitters.}
\label{app:TwoEmitters}
The calculation presented in Sec. \ref{sec:Emission From Coherent Superposition} can be done for the case of 2 emitters as well, leading to similar conclusions. Here, the most general pure-state initial condition can be written as
\begin{equation}
\label{eq:superpositionTwoEmitters}
 \ket{\psi(0)}=\alpha_0\ket{1,-1}+\beta_0\ket{1,0}+\gamma_0\ket{1,1}.
\end{equation}
Then, assuming no jump between time $0$ and $t$, the state at time $t$ can be written as:
\begin{equation}
 \ket{\psi(t)}=\frac{\alpha_0\ket{1,-1}+\me^{-\Gamma t}\left(\beta_0\ket{1,0}+\gamma_0\ket{1,1}\right)}{\sqrt{|\alpha_0|^2+\me^{-2\Gamma t}\left(|\beta_0|^2+|\gamma_0|^2\right)}}.
\end{equation}
From this, the decay rate conditioned on that no jump has yet occurred can be calculated in a very similar way to \eqref{eq:Jump rate from superposition}, to obtain
\begin{multline}
r(t)=\Gamma\,\bra{\psi \left( t \right)} \hat{J}^+ \hat{J}^- \ket{\psi \left( t \right)}\\=
\frac{2\Gamma\,\me^{-2\Gamma t}\left(|\beta_0|^2+|\gamma_0|^2\right)}{\left| \alpha_0 \right|^2 + \me^{-2\Gamma t}\left(|\beta_0|^2+|\gamma_0|^2\right)}.
\end{multline}
Then, the probability of no jump occurring before time $t$ reads:
\begin{multline}
p \left\{\text{no jump before } t \right\} = \exp\left(-\int_0^td\tau\;r(\tau)\right) \\= |\alpha_0|^2 + \me^{-2\Gamma t}\left(|\beta_0|^2+|\gamma_0|^2\right).
\end{multline}
From this, we can immediately read off that with a probability $|\alpha_0|^2$, no jump will ever occur. On the other hand, if the first jump occurs, then the state becomes
\begin{equation}
 \ket{\psi}_\text{after 1st jump}=\frac{\beta_0\ket{1,-1}+\gamma_0\ket{1,0}}{\sqrt{|\beta_0|^2+|\gamma_0|^2}},
\end{equation}
independently of when the jump occurs. From this point on, the dynamics is the same as with a single emitter, only with a doubled decay rate $2\Gamma$. That is, no more jump will occur with a probability $|\beta_0|^2$.

In summary, 0 jump will occur with a probability $|\alpha_0|^2$, 1 with a probability $|\beta_0|^2$, and 2 with $|\gamma_0|^2$. In every other respect as well, the behavior will be the same as if the system were started from the mixture 
\begin{multline}
 \hat{\rho}(0)=|\alpha_0|^2\ket{1,-1}\bra{1,-1}\\+|\beta_0|^2\ket{1,0}\bra{1,0}+|\gamma_0|^2\ket{1,1}\bra{1,1}
\end{multline}
instead of the superposition \eqref{eq:superpositionTwoEmitters}.

\end{document}